\documentclass[aps,nofootinbib,prd,eqsecnum,showpacs,showkeys,preprintnumbers,altaffilletter]{revtex4-1}

\usepackage[normalem]{ulem}
\usepackage{amsmath}
\usepackage{amssymb}

\usepackage{graphicx}
\usepackage{amsfonts}
\usepackage{color}
\usepackage{bm}
\usepackage{mathrsfs}
\usepackage{epstopdf}
\usepackage{url}
\usepackage{footnote}
\usepackage{appendix}

\usepackage{float}
\usepackage{enumerate}
\usepackage{lineno}

\usepackage{hyperref}

\usepackage{tabularx}

\makeatletter

\newcommand{\stkout}[1]{\ifmmode\text{\sout{\ensuremath{#1}}}\else\sout{#1}\fi}

\usepackage{booktabs}
\AtBeginDocument{
\heavyrulewidth=.08em
\lightrulewidth=.05em
\cmidrulewidth=.03em
\belowrulesep=.65ex
\belowbottomsep=0pt
\aboverulesep=.4ex
\abovetopsep=0pt
\cmidrulesep=\doublerulesep
\cmidrulekern=.5em
\defaultaddspace=.5em
}

\newcolumntype{L}[1]{>{\hsize=#1\hsize\raggedright\arraybackslash}X}%
\newcolumntype{R}[1]{>{\hsize=#1\hsize\raggedleft\arraybackslash}X}%
\newcolumntype{C}[1]{>{\hsize=#1\hsize\centering\arraybackslash}X}%

\newcommand*\patchAmsMathEnvironmentForLineno[1]{%
 \expandafter\let\csname old#1\expandafter\endcsname\csname #1\endcsname
 \expandafter\let\csname oldend#1\expandafter\endcsname\csname end#1\endcsname
 \renewenvironment{#1}%
   {\linenomath\csname old#1\endcsname}%
   {\csname oldend#1\endcsname\endlinenomath}}%
\newcommand*\patchBothAmsMathEnvironmentsForLineno[1]{%
 \patchAmsMathEnvironmentForLineno{#1}%
 \patchAmsMathEnvironmentForLineno{#1*}}%
\AtBeginDocument{%
\patchBothAmsMathEnvironmentsForLineno{equation}%
\patchBothAmsMathEnvironmentsForLineno{align}%
\patchBothAmsMathEnvironmentsForLineno{flalign}%
\patchBothAmsMathEnvironmentsForLineno{alignat}%
\patchBothAmsMathEnvironmentsForLineno{gather}%
\patchBothAmsMathEnvironmentsForLineno{multline}%
}

\definecolor{greenNew}{rgb}{0.15,0.75,0.0}

\makeatother

\begin{document}

\title{ A varying Dark Energy effective speed of sound parameter in the phantom Universe}

\author{Imanol Albarran $^{1,2}$}
\email{imanol@ubi.pt}

\author{Mariam Bouhmadi-L\'{o}pez $^{3,4}$}
\email{{\mbox{mariam.bouhmadi@ehu.eus}}}

\author{Jo\~{a}o Marto $^{1,2}$}
\email{{\mbox{jmarto@ubi.pt}}}


\date{\today}

\affiliation{
${}^1$Departamento de F\'{\i}sica, Universidade da Beira Interior, Rua Marqu\^{e}s D'\'Avila e Bolama 6200-001 Covilh\~a, Portugal\\
${}^2$Centro de Matem\'atica e Aplica\c{c}\~oes da Universidade da Beira Interior, Rua Marqu\^{e}s D'\'Avila e Bolama 6200-001 Covilh\~a, Portugal\\
${}^3$Department of Physics,  University of the Basque Country UPV/EHU. P.O. Box 644, 48080 Bilbao, Spain\\
${}^4$IKERBASQUE, Basque Foundation for Science, 48011, Bilbao, Spain\\
}

\begin{abstract}

We analyse the phenomenological effects of a varying Dark Energy (DE) effective speed of sound parameter, $c^{2}_{\textrm{sd}}$, on the cosmological perturbations of three phantom DE models. Each of these models induce a particular abrupt future event known as Big Rip (BR), Little Rip (LR), and Little Sibling of the Big Rip (LSBR). In this class of abrupt events, all the bound structures in the Universe would be ripped apart at a finite cosmic time.  We compute the evolution of the perturbations, $f\sigma_{8}$ growth rate and forecast the current matter power spectrum. We vary the $c^{2}_{\textrm{sd}}$ parameter in the interval $[0,1]$ and compute the relative deviation with respect $c^{2}_{\textrm{sd}}=1$. In addition, we analyse the effect of gravitational potential sign flip that occurs at very large scale factors as compared with the current one.
\end{abstract}



\maketitle

%
%

During the last two decades Cosmology has experienced a great improvement  in the theoretical and observational scopes. The discovery of an accelerated Universe, a fact supported by several observations \cite{Riess:1998cb,Perlmutter:1998np}, has developed a flourishing of new ideas that deal with the intriguing current speed up. The simplest explanation consists into invoking a new component in the Universe named DE as the responsible of the current acceleration \cite{AmendolaTsujikawa}. Among the vast amount of DE models, those where the null energy condition is violated are coined as phantom \cite{Caldwell:2003vq,Caldwell:1999ew,Dabrowski:2003jm,Stefancic:2003rc}. In these class of models, the Equation of State (EOS) parameter of DE, $w_{\textrm{d}}$ (the ratio between pressure and energy density of DE), stays always below $-1$. Despite some energy conditions are not satisfied, phantom DE models seem to be favoured by observations \cite{DiValentino:2016hlg,Vagnozzi:2018jhn,Vagnozzi:2019ezj,Alestas:2020mvb,DiValentino:2020vnx,DiValentino:2020naf,Yang:2021hxg,Yang:2021egn,DiValentino:2021izs,Aghanim:2018eyx}.

\

It is known that most phantom DE models predict future singularities. As we have already mentioned in the introduction section, we focus on three genuine phantom models where each of them induce a particular future doomsday known as BR, LR, and LSBR ( see \cite{Caldwell:2003vq,Caldwell:1999ew,Dabrowski:2003jm,Starobinsky:1999yw,Carroll:2003st,Chimento:2003qy,GonzalezDiaz:2003rf,GonzalezDiaz:2004vq}, \cite{Ruzmaikina,Bouhmadi-Lopez:2013nma,Nojiri:2005sx,Nojiri:2005sr,Stefancic:2004kb,BouhmadiLopez:2005gk,Frampton:2011sp,Brevik:2011mm}, \cite{Bouhmadi-Lopez:2014cca,Bouhmadi-Lopez:2018lly,Morais:2016bev} for a detailed description of the respective models). We recall that no matters if a true singularity or an abrupt event takes place, all the bound structures in the Universe are torn away and destroyed.

\

All the models mentioned above can be understood as alternatives to the widely accepted $\Lambda$CDM paradigm, and therefore, good models to describe suitably the current Universe. An appropriate fitting of the parameters involved could make these models indistinguishable among them at the background level. Therefore, it becomes necessary to address the cosmological perturbations as well. 

\

Observables as for example, the matter power spectrum and the growth rate  provide useful data about the distribution of matter. Unfortunately,  in most of the cases the imprints of different DE models on such observables are insignificant. Therefore, important efforts have been made to improve the accuracy of the observations, particularly, focusing on scrutinising the DE sector as it is the case of Euclid mission \cite{Laureijs:2011gra,Amendola:2016saw}. 

\

The squared speed of sound parameter, ${c}_{s}^2$, is another important variable that plays a key role on cosmological perturbations. It is well known that DE models with a negative ${c}_{s}^2$ parameter induce instabilities at the perturbation level. To avoid those instabilities, in \cite{Bean:2003fb,Valiviita:2008iv} the authors consider a non-adiabatic  contribution on the pressure perturbations. This method lead to separate the adiabatic speed of sound, $c_a$;  which depends on the EoS,  and the rest frame speed of sound (often coined as the effective speed of sound), $c^{2}_s$; which is regarded as a free parameter\footnote{In the case of a scalar field representation, the effective speed of sound parameter coincides with unity, i.e. $c_s^2=1$ (c.f please \cite{Bean:2003fb,Valiviita:2008iv} for a detailed explanation).}. Several works have addressed the issue of pressure decoupling in different DE models. For example, in \cite{DeDeo:2003te} the authors study the implications of a time varying speed of sound in quintessence models. In \cite{ul:2014dxa} the authors analyse the effects of a varying effective speed of sound parameter on the matter perturbations for a DE content described by a scalar field. A recently introduced pretty interesting approach consists on modelling the effective speed of sound as a function of the EoS parameter i.e.  ${c}_{s}^2= {c}_{s}^2 (w)$. For example, in \cite{Perkovic:2020eju} the authors show how to reconstruct the Lagrangian (in particular, for those models with a purely kinetic term) starting from a known ${c}_{s}^2 (w)$ function and address several approaches for a range of different DE models. In particular, the so called \emph{effective field theory of DE} consists on a Lagrangian description of the cosmological perturbations \cite{Frusciante:2019xia}. The effective speed of sound parameter will be given by the fundamental Lagrangian, therefore,  this formalism could be potentially efficient to check the validity of DE  models and fitting the DE speed of sound parameter. In \cite{Nesseris:2015fqa}, the authors consider a model with a constant EoS parameter and estimate the corrections on the growth index when changing $c_{s}^2$. On the other hand, in \cite{Pietrobon:2008js} it is considered a DE model with an affine EoS. Then, the results obtained when fixing $c_{s}^2=0$ and $c_{s}^2=1$ are compared. A further analysis on the effective speed of sound parameter is performed in \cite{Ballesteros:2010ks}, where the authors consider the contribution of matter (Baryonic and dark matter (DM)), photons and neutrinos to get, for example, a probability distribution for the $c_{s}^2$ value. In \cite{Linton:2017ged}, a new class  of DM-DE interacting models is identified. The authors study the implications of a varying effective speed of sound on the Cosmic Microwave Background (CMB) and the matter power spectrum. 

\

 An interesting extension of the widely studied DE models consists in considering the effects of anisotropy and viscosity. In \cite{Koivisto:2005mm} the authors address the effects of viscosity on the CMB and matter power spectrum for a Generalised Chaplying Gas and models with a constant EoS parameter (both standard and phantom type DE matter). Then, the obtained results are compared against the effects that a non-vanishing effective speed of sound could induce. Furthermore, in \cite{Arora:2020lsr} the authors analyse the effects of a viscosity bulk within a modified gravity scenario endowed with the general action $f(R,T)$ and perform a test to check the validity of the studied models. The impact of  the non adiabatic perturbations have been studied for example in \cite{Velten:2017mtr,Arjona:2020yum}. In \cite{Velten:2017mtr}, the authors address a particular parametrisation of DE considering a linear combination of the intrinsic and entropy perturbations. In the recent work \cite{Arjona:2020yum}, aside the non-adiabatic perturbations  the non-vanishing anisotropic stress tensor is regarded as well. Here, the authors use both analytical and numerical solutions of the growth rate to compare with the latest observational data. On the other hand, in the recent work \cite{Arjona:2020kco}, the authors use machine learning computation methods to reconstruct the  relevant perturbation parameters including those involved with the anisotropic effects, pointing out a way to detect imprints of anisotropies on a wide range of DE models. As it is shown in \cite{Koivisto:2005mm,Arjona:2020kco}, when considering such anisotropies the DE sound speed could be negative without inducing instabilities at the perturbation level as long as the effective speed of sound parameter stands positive.

\

A model that has gained some attention recently is the so called Early Dark Energy (EDE) model, which has been shown to be slightly favoured by observational data. This model simply consists on considering a small but not negligible DE presence at early stages of the Universe (for example, before the matter-radiation decoupling time) which could induce significant footprints on the structure formation \cite{Niedermann:2019olb,Chudaykin:2020acu}. Following this new research line the relevant model parameters were observationally constrained together with the  $c_{s}^2$ parameter in \cite{Bhattacharyya:2019lvg} and considering the additional viscosity speed of sound parameter in \cite{Verde:2016wmz}.  Other  models have been  observationally constrained in order to fit a value for $c_{s}^2$. For example, in \cite{Bean:2003fb,dePutter:2010vy} the authors use the temperature fluctuations of the CMB dataset to set the value of the speed of sound. In \cite{Hu:2004yd} the authors measure the effects of DE clustering on the large scale structure using CMB and the galaxy clustering cross correlation data.  On the other hand, in \cite{Takada:2006xs} the author analyse the effects of DE clustering on the structure formation at large scales and forecast the upper bounds on the DE speed of sound parameter to distinguish among DE models. In \cite{Mehrabi:2015hva}, the authors study the structure formation and constrain a CPL model with a free effective speed of sound parameter. Significant results could be obtained using the large neutral Hydrogen surveys as it has been shown in \cite{TorresRodriguez:2007mk,TorresRodriguez:2008et}, where the authors highlight the potential of the Square Kilometre Array to constrain DE models. In view of the upcoming Euclid mission, several works forecast the necessary accuracy in order to discriminate between different DE models. For instance, in \cite{Sapone:2013wda} the authors compute the sensitivity of the photometric and spectroscopic surveys for measuring the speed of sound and viscosity parameters. 

\

In this work, we consider a Universe filled with radiation, matter and DE components, where the latter is described by three different phantom models. We have focused our attention on the models coined, in the present work, as model A, model B and model C since they share the common feature that their induced singularities and abrupt events (BR, LR and LSBR respectively) are genuinely phantom, that is, they only occur if and only if a phantom type of component is present. We address the scalar cosmological perturbations following the method of pressure decomposition for DE \cite{Bean:2003fb,Valiviita:2008iv}. We set the initial conditions as done in \cite{Morais:2015ooa,Albarran:2016mdu} where the physical value  of the total matter density contrast, $\delta_{{\rm phys.}}(k)$, for a single field inflation is taken from Planck data 2018 \cite{wikiesa}. On the other hand, the background parameters are fixed following \cite{Bouali:2019whr}.  After imposing adiabatical conditions for scales larger than the horizon at the beginning, the physical value  of  $\delta_{{\rm phys.}}(k)$ is the last condition needed to ultimately fix all the initial numerical values.  We analyse the phenomenological effects of changing the effective speed of sound on the perturbations. We emphasise that a complete constraint of the parameters should involve  both, the background parameters and the $c^{2}_{s\textrm{d}}$ parameter at the perturbation level. However, we are not interested in constraining $c^{2}_{s\textrm{d}}$ but in the phenomenological effect of varying it. Therefore, we set the background parameters following \cite{Bouali:2019whr} and  then, we compute the perturbations considering, as the simplest choice, a constant value of $c^{2}_{s\textrm{d}}$ inside the interval $\left[0,1\right]$. In addition, we compute the relative differences on observables by evaluating the matter power spectrum and $f\sigma_8$ growth rate. Finally, we study the behaviour of the gravitational potential on large scales and large scale factors as compared with current one.

\

The paper is organised as follows, In section \ref{review} we briefly review the background of the models inducing the BR, LR, and LSBR events. In section~\ref{results_Speed} we present the obtained results and in section~\ref{conclusions_Speed}, we present the main conclusions. Finally, in the appendix  \ref{pressure_decompose} we show in detail the pressure decomposition into its adiabatic and non-adiabatic contributions.

%
%
%
%

\section{Background models}
\label{review}

In this section, we introduce the background of three genuine phantom DE models. Each of this models induce a particular abrupt event known as;   Big Rip (model A), Little Rip (model B) and Little Sibling of the Big Rip (model C).  We start by considering an isotropic and homogeneous Universe, where the geometry  is given by the Friedman-Lema\^{i}tre-Robertson-Walker (FLRW) space-time metric:
\begin{equation}\label{metric}
ds^{2}=-dt^{2}+a^{2}\left(t\right)\left[dx^{2}+dy^{2}+dz^{2}\right].
\end{equation}
We have considered the case of a spatially flat Universe  in agreement with observations \cite{Aghanim:2018eyx}. Therefore, the  Friedmann and Raychaudhuri equations are written as follows
\begin{eqnarray}
\label{Friedman}
	H^2=\frac{8\pi G}{3}\rho
	\,,
\end{eqnarray}
\begin{eqnarray}
\label{conservback}
	\dot{H}=-4\pi G\left(\rho +p\right)
	\,,
\end{eqnarray}
where $G$ is the gravitational constant, $\rho$ is the total energy density of the Universe, while $p$ is likewise the total pressure. We assume that each component is independently conserved, therefore, the conservation equation reads
\begin{eqnarray}
\label{conservindiv}
	\dot{\rho}_{\ell} + 3H\left(\rho_{\ell}+p_{\ell}\right)=0,
\end{eqnarray}
where, $\ell=\textrm{r},\textrm{m},\textrm{d}$ stands for radiation, matter and DE, respectively. In consequence, the Friedman equation can be written as
\begin{equation}
\begin{split}
H^{2}=H_{0}^{2}\Bigl[\Omega_{\textrm{r}0}a^{-3\left(1+w_{\textrm{r}}\right)}+
      \Omega_{\textrm{m}0} a^{-3\left(1+w_{\textrm{m}}\right)}
      +\Omega_{\textrm{d}0}f_{\textrm{j}}\left(a\right) \Bigr]
\end{split}\label{developedFriedman}
\end{equation}
where $H$ is the Hubble parameter, $a$ is the scale factor and the parameters $\Omega_{\ell 0}$ ($\ell=\textrm{r},\textrm{m},\textrm{d}$) are the current fractional energy densities of the aforementioned components. The subindex $0$ denotes the values at present time. From now on, we will adopt $a_{0}=1$. In order to avoid repetitions on the notation, the scale factor will be denoted simply by $a$. While the EoS parameters for radiation ($w_{\textrm{r}}=1/3$) and matter ($w_{\textrm{m}}=0$) are constant, it can be scale dependent in the case of DE. The contribution of DE to the total energy budget can be  expressed by means of the dimensionless function $f_{\textrm{j}}\left(a\right)$, where the subindex j refers to the selected model (j=A,B,C). 

\ 

The  set of parameters corresponding to each models are fixed  by using the constraints obtained in the work \cite{Bouali:2019whr}. The necessary parameters to totally describe the background models are: The current fractional energy densities of radiation and matter, $\Omega_{\textrm{r}0}$ and $\Omega_{\textrm{m}0}$; the current Hubble parameter, $H_0$ (cf. table III in \cite{Bouali:2019whr}). While to get the perturbations we need: The root mean square mass fluctuations amplitude in spheres of size $8$ $\textrm{h}^{-1}\textrm{Mpc}$, $\sigma_8$; the amplitude of the scalar perturbations as predicted for single field inflation, $A_s$,  and the spectral index, $n_s$. These later parameters are fixed by Planck data (cf. table 17.18 in \cite{wikiesa}).

\

We understand that the differences between background models mostly lie on the $f_{\textrm{j}}\left(a\right)$ function, while we expect to find footprints of different DE models, (i) at present, in such a way that  they could be useful to distinguish between different DE models, and (ii), in the far future, where such deviations between DE models become larger and enhance some features of each particular DE model.

%
%

\subsection{model A: BR singularity}
\label{BR}
The BR singularity can be induced by a DE content characterised by the following EoS \cite{Caldwell:2003vq,Caldwell:1999ew,Dabrowski:2003jm,Starobinsky:1999yw,Carroll:2003st, Chimento:2003qy,GonzalezDiaz:2003rf,GonzalezDiaz:2004vq},
\begin{equation}\label{EoSmodelA}
p_{\textrm{d}}=w_{\textrm{dA}}\rho_{\textrm{d}},
\end{equation}
where $w_{\textrm{dA}}$ is  a constant and  smaller than $-1$. Solving the conservation equation we get the expression for the corresponding $f_{\textrm{A}}\left(a\right)$ function in Eq. (\ref{developedFriedman})
\begin{equation}\label{famodelA}
f_{\textrm{A}}\left(a\right)=a^{-3\left(1+w_{\textrm{dA}}\right)}.
\end{equation}
%

%
%

\subsection{model B: LR abrupt event}
\label{LR}
The model B induces a LR abrupt event \cite{Ruzmaikina,Bouhmadi-Lopez:2013nma,Nojiri:2005sx,Nojiri:2005sr,Stefancic:2004kb,BouhmadiLopez:2005gk,Frampton:2011sp,Brevik:2011mm}  and can be identified by having the following EoS for DE content \cite{Nojiri:2005sx,Stefancic:2004kb}
\begin{equation}\label{EoSmodelB}
p_{\textrm{d}}=-\rho_{\textrm{d}}-\mathcal{B}\sqrt{\rho_{\textrm{d}}},
\end{equation}
where $\mathcal{B}$ is a positive constant which has the dimension of an inverse squared length. This model can be understood as a deviation of the widely known $\Lambda$CDM paradigm. Notice that for a vanishing parameter $\mathcal{B}$ the $\Lambda$CDM model is recovered.  Solving the conservation equation we get the corresponding $f_{\textrm{B}}\left(a\right)$ function for model B \cite{Ruzmaikina,Bouhmadi-Lopez:2013nma,Nojiri:2005sx,Nojiri:2005sr,Stefancic:2004kb,BouhmadiLopez:2005gk,Frampton:2011sp,Brevik:2011mm},
\begin{equation}\label{famodelB}
f_{\textrm{B}}\left(a\right)=\left[1+\frac{3}{2}\sqrt{\frac{\Omega_{\textrm{B}}}{\Omega_{\textrm{d}0}}}\ln\left(a\right)\right]^2,
\end{equation}
where the parameter $\mathcal{B}$ is reabsorbed in the dimensionless parameter $\Omega_{\textrm{B}}\equiv\left[\left(8\pi G\right)/\left(3H_{0}^{2}\right)\right]\mathcal{B}^{2}$.  This class of abrupt event suffers from all the divergences prevalent in a BR singularity but driven at infinite cosmic time. Therefore, we consider a LR less harming that a BR. 
%
%
%

\subsection{model C: LSBR abrupt event}
\label{LSBR}
This model induces a LSBR abrupt eventy and it is distinguished by having the following EoS \cite{Bouhmadi-Lopez:2014cca,Bouhmadi-Lopez:2018lly,Morais:2016bev}
\begin{equation}\label{EoSmodelC}
p_{\textrm{d}}=-\rho_{\textrm{d}}-\frac{\mathcal{C}}{3},
\end{equation}
where $\mathcal{C}$ is a positive constant. The smaller is $\mathcal{C}$, the closer is the model C to $\Lambda$CDM. Solving  the conservation equation we get the corresponding expression of $f_{\textrm{C}}\left(a\right)$ for model C \cite{Bouhmadi-Lopez:2014cca},
\begin{equation}\label{famodelC}
f_{\textrm{C}}\left(a\right)=1+\frac{\Omega_{\textrm{C}}}{\Omega_{\textrm{d}0}}\ln\left(a\right),
\end{equation}
where the constant $\mathcal{C}$ is absorbed in the new parameter $\Omega_{\textrm{C}}\equiv\left[\left(8\pi G\right)/\left(3H_{0}^{2}\right)\right]\mathcal{C}$. The model C induces the abrupt event known as LSBR. In this kind of abrupt event, the scale factor and the Hubble parameter diverge at infinite cosmic time while the first cosmic time derivative of the Hubble parameter is finite. We regard the  LSBR   as the less harming abrupt event among those induced by phantom scenarios.


\section{Results: The effect of the speed of sound}
\label{results_Speed}
In the following, we present the results obtained for the  cosmological  perturbations evolution and for the three models addressed in this paper. We remind that in order to set the model parameters we have used those obtained in our previous work \cite{Bouali:2019whr}. We compute the evolution of the matter density contrast and peculiar velocities, from well inside the radiation dominated epoch\footnote{The scale factor for this epoch represents a moment in the early Universe where  its energy content consists in $1\%$ of matter against $99\%$ of radiation.}, $a\sim2.65\times 10^{-6}$, till the far future, $a\sim1.62\times10^{5}$. We perform the integrations for the following six particular modes
 \begin{itemize}
 \item small k (large distances): $k_{1}=3.33\times10^{-4}\textrm{h Mpc}^{-1} $ and $k_{2}=1.04\times10^{-4}\textrm{h Mpc}^{-1}$.
 \item medium k (intermediate distances): $k_{3}=3.26\times10^{-3}\textrm{h Mpc}^{-1}$ and $k_{4}=1.02\times10^{-2}\textrm{h Mpc}^{-1}$.
 \item  large k (short distances): $k_{5}=3.19\times10^{-2}\textrm{h Mpc}^{-1}$ and $k_{6}=1.00\times10^{-1}\textrm{h Mpc}^{-1}$.
 \end{itemize}
The minimum mode, $k_{1}$, coincides with the current Hubble horizon, i.e.  no smaller mode can be detected. On the other hand,  we consider as maximum mode, $k_{6}$, where the linear approximation breaks down and the  non-linear contributions become important.

\subsection{Matter power spectrum and $f\sigma_{8}$}

 We have computed the current matter power spectrum and the growth rate $f\sigma_{8}$,  testing the effective squared speed of sound  from $0$ to $1$ in steps of $0.2$. In  this process, the numerical integration was repeated for 200 modes ranged from $k_1$ to $k_6$. 
\begin{figure*}[h!]
 \includegraphics[width=\textwidth]{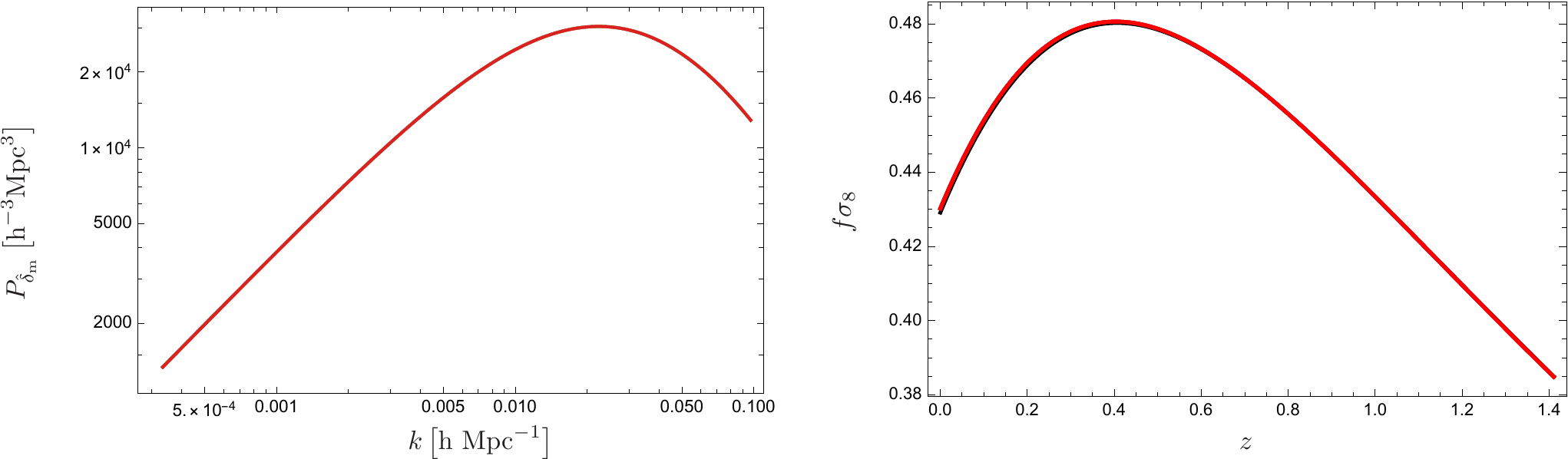}
\caption{The left panel of this figure  represents the matter power spectrum while the right panel shows the evolution of $f\sigma_{8}$ in terms of the redshift $z$. All models with different values of $c^{2}_{\textrm{sd}}$ give an almost identical result, so the curves appear completely overlapped and their differences are negligible.}\label{genmpsfs8}
\end{figure*}
\begin{figure*}[h!]
 \includegraphics[width=15.9cm]{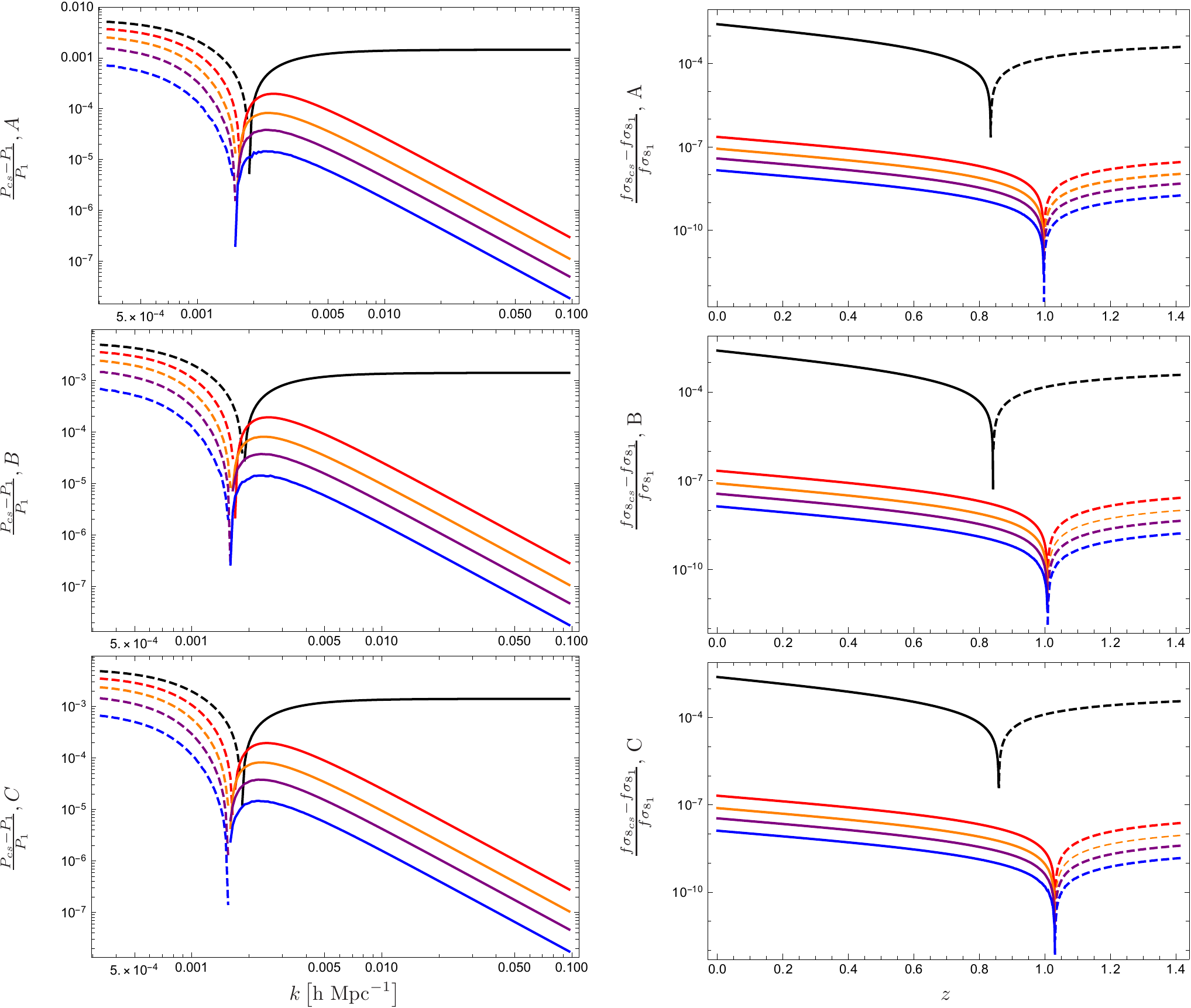}
\caption{These plots represent the relative deviation  with respect to the result given when $c^{2}_{s\textrm{d}}=1$. The top, middle and bottom panels correspond with the models A, B and C, respectively. The left panels show the results for the matter power spectrum  as a function of the mode. The right panels show the results for  $f\sigma_8$ in terms of redshift, $z$. Different values of $c^{2}_{s\textrm{d}}$ are coloured as; $c^{2}_{s\textrm{d}}=0$ (black), $c^{2}_{s\textrm{d}}=0.2$ (red), $c^{2}_{s\textrm{d}}=0.4$ (orange), $c^{2}_{s\textrm{d}}=0.6$ (purple) and $c^{2}_{s\textrm{d}}=0.8$ (blue). The plots are represented in a logarithmic scale,  in such a way that dashed lines correspond with negative values while solid lines represent positive values.}\label{mpsrel}
\end{figure*}

\

Figure~\ref{genmpsfs8} shows  the current matter power spectrum and the evolution of $f\sigma_{8}$ predicted by the models. These results are in agreement with observations but does not allow to distinguish any deviation for different models. In addition,  the effects of a varying speed of sound turn out to be almost undetectable since the results appear totally overlapped. Therefore, in order to give an account of the contrast, we compute the relative deviation with respect to $c^{2}_{s\textrm{d}}=1$. 

\

As it is shown in the left column of figure~\ref{mpsrel}, the relative differences (with respect to $c^{2}_{s\textrm{d}}=1$) on  the matter power spectrum are negative for the smallest modes and positive for the largest ones. The transition occurs in a narrow interval around the wave number $k\sim1.8\times10^{-3} \textrm{h} \ \textrm{Mpc}^{-1}$.  The separation obtained for a vanishing speed of sound parameter is remarkable. First, looking at small modes, the deviations are constant, the larger is the deviation from $c^{2}_{s\textrm{d}}=1$ the larger is such a constant. Secondly, looking at the larger modes, the deviation is constant for vanishing $c^{2}_{s\textrm{d}}$ parameters while such deviation vanishes for non vanishing $c^{2}_{s\textrm{d}}$ parameters. 

\

Something similar happen for $f\sigma_8$ results. As it is shown on the right panel of figure~\ref{mpsrel}, the relative difference for a vanishing effective speed of sound  parameter show an important separation with respect to the results given  for  a non-vanishing $c^{2}_{s\textrm{d}}$ parameter.  Conversely, the deviation is positive for the smallest redshifts and negative for the largest ones. The transition occurs at $z\sim0.85$ for a vanishing 
$c^{2}_{s\textrm{d}}$ and at $z\sim1$ for non-vanishing $c^{2}_{s\textrm{d}}$. Such transition point is slightly affected depending which DE model is considered. In addition, contrary to what happens for the matter power spectrum, in $f\sigma_8$  the deviations goes to a  constant for both large and small modes. As expected, such a constant is larger the larger is the deviation from $c^{2}_{s\textrm{d}}=1$.

\

The largest deviations are of the order $10^{-3}$ for both the matter power spectrum and $f\sigma_8$ evolution. So we conclude that no significant footprints  appear on the matter distribution when changing the effective speed of sound. In fact, the most relevant effects of a varying effective speed of sound are clearly manifested in the DE sector.

\subsection{DE perturbations}

\begin{figure*}[h!]
 \includegraphics[width=\textwidth]{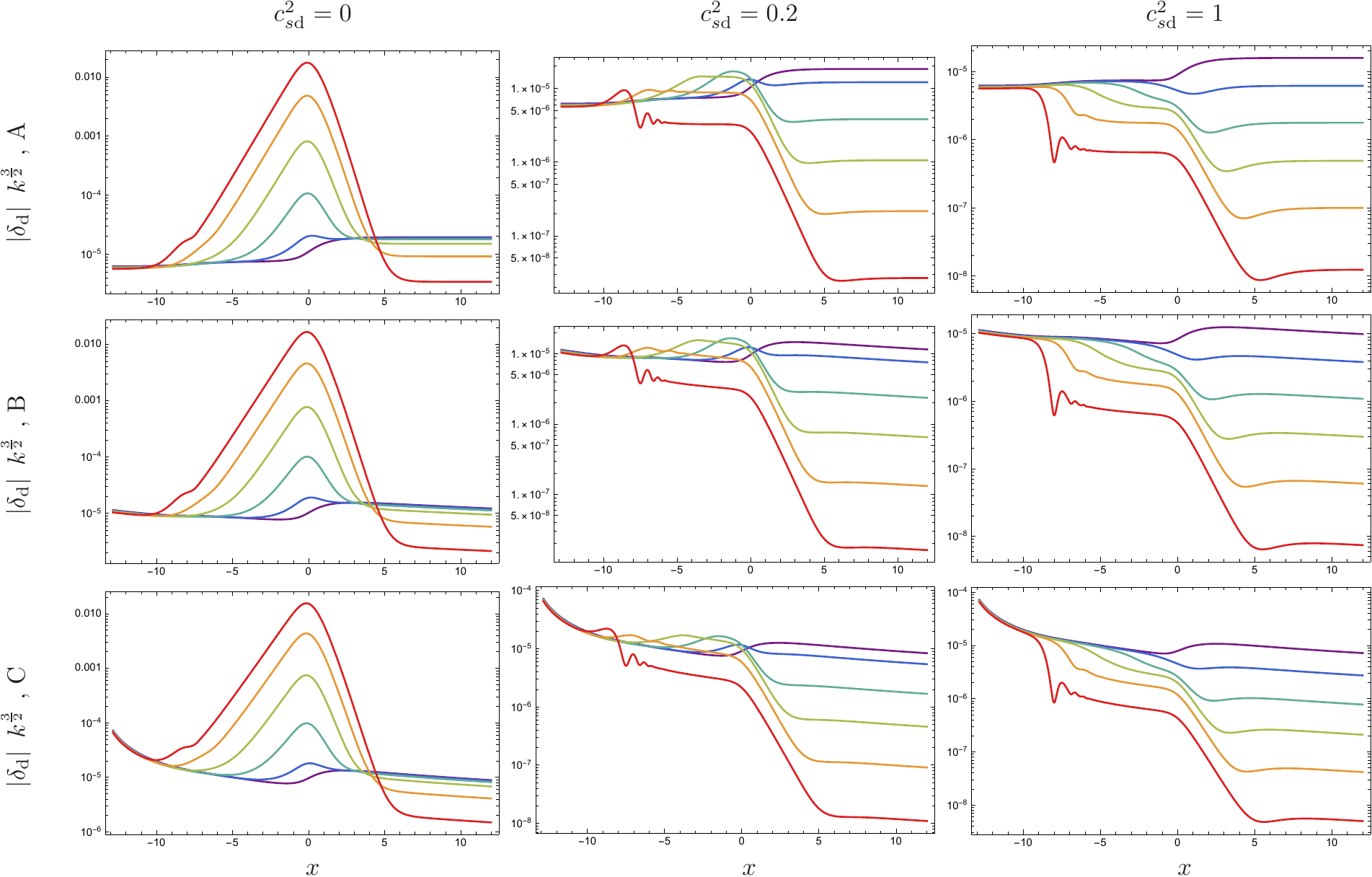}
\caption{These plots show the evolution of DE density contrast  for different models and different values of $c_{s\textrm{d}}^{2}$. The panels of the first, second and third row correspond with the model A, B and C, respectively. The panels of the first, second and third columns correspond, respectively, to the values of the squared speed of sound $c_{s\textrm{d}}^{2}=0$, $c_{s\textrm{d}}^{2}=0.2$ and $c_{s\textrm{d}}^{2}=1$. The plot is drawn as a function of $x=\ln\left(a\right)$ which goes from well inside the radiation dominated epoch,  $x^{\star}=-12.84$, to the far future, $x=12$. Each colour corresponds to a particular value of the wave-number $k$: $k_1=3.33\times10^{-4}\textrm{h Mpc}^{-1}$ (purple), $k_{2}=1.04\times10^{-4}\textrm{h Mpc}^{-1}$ (dark blue), $k_{3}=3.26\times10^{-3}\textrm{h Mpc}^{-1}$ (light blue), $k_{4}=1.02\times10^{-2}\textrm{h Mpc}^{-1}$ (green), $k_{5}=3.19\times10^{-2}\textrm{h Mpc}^{-1}$ (orange)  and $k_{6}=1.00\times10^{-1}\textrm{h Mpc}^{-1}$ (red).}\label{deltadall}
\end{figure*}

 Figure~\ref{deltadall} shows the evolution of the matter density contrast of DE for different models and ranges of  $c^{2}_{s\textrm{d}}$. We remind that due to the phantom nature of DE models, the adiabatic condition imposed at the early Universe  implies that the DE perturbations are negative \cite{Albarran:2016mdu,Albarran:2017kzf}.

\

Figure~\ref{deltadall} shows the evolution of DE density contrast for different models and values of $c_{s\textrm{d}}^{2}$. We realise that the effect of varying  $c^{2}_{s\textrm{d}}$ is minimal for values larger than $0.2$, i.e., inside the interval $\left[0.2, 1\right]$. Hence,  we  just show the results for $c_{s\textrm{d}}^{2}=0$, $c_{s\textrm{d}}^{2}=0.2$ and $c_{s\textrm{d}}^{2}=1$. As can be seen in the first column of Figure~\ref{deltadall} (i.e. for a  vanishing $c^{2}_{s\textrm{d}}$ parameter), once the modes enters  the horizon, the perturbations increase up to three orders of magnitude in the case of the largest mode and around two orders of magnitude for the medium size modes. All the growing modes reach a maximum at present time ($x=0$, where $x=\ln\left(a\right)$) and decay during the later DE domination era. That is;  once the modes enters the horizon they grow, and then, when they exit the horizon, the perturbations decay evolving towards a negative constant\footnote{We remind that phantom DE perturbations are considered to be negative at the beginning of the numerical integration,  as it is the case of the gravitational potential. On the other hand, standard matter perturbations are considered to be positive.}. This is not the case of the smallest modes, we should bear in mind that such small modes have recently entered the horizon and are the first exiting it, so the smallest modes do not experience important deviations.
  
\

For a $c^{2}_{s\textrm{d}}=0.2$, the growth of DE perturbations is strongly suppressed in the matter domination era. During this epoch, the largest modes  ($k_5$ and $k_6$) decay and reach a plateau while the medium sized modes ($k_3$ and $k_4$) experience a small growth.  When  DE starts dominating,  the perturbations decrease up to three orders of magnitude for the largest modes and one order of magnitude for the medium sized modes. The smallest modes ($k_1$ and $k_2$) do not seem to be significantly affected.

\

For a value of  $c^{2}_{s\textrm{d}}=1$, the resulting plot is very similar to the one when  $c^{2}_{s\textrm{d}}=0.2$. The main difference consists on the total suppression of the growing perturbations during the matter dominated epoch.  Once again, the perturbations decay when the corresponding mode enters the horizon and evolve to a negative constant after exiting the horizon.

\

In summary, DE perturbation are strongly affected near vanishing values of $c^{2}_{s\textrm{d}}$ parameter and mostly, for large modes. On the contrary, small modes do not show significant deviations. We should bear in mind that due to the change of the acceleration of the Universe (from a negative to a positive acceleration stage) the smallest modes are the last entering the horizon and the first exiting it, therefore, such modes have not enough time to be significantly affected. 

\

On the other hand, it is possible to find important deviations between the different models, mostly, in the early Universe where radiation dominates over the other components. We set the initial value of DE matter density contrast, $\delta_{\textrm{d}}^{\star}$ (where the script $^{\star}$ denotes the initial value) through the adiabatic condition \cite{AmendolaTsujikawa}.

\begin{eqnarray}\label{initialdeltaeq}
\frac{\delta^{\star}_{\textrm{r}}}{1+w^{\star}_{\textrm{r}}}=\frac{\delta^{\star}_{\textrm{m}}}{1+w^{\star}_{\textrm{m}}}=\frac{\delta^{\star}_{\textrm{d}}}{1+w^{\star}_{\textrm{d}}}.
\end{eqnarray}

Taking into account that we have used in all the models the same value of the current radiation fractional energy, $\Omega_{\textrm{r}0}$, and that the current matter fractional energy is almost the same in the three paradigms analysed, $\Omega_{\textrm{m}0}\simeq0.3$, it is worthy to point out the next approximation relating the initial DE perturbations of the different models
\begin{eqnarray}\label{initialdeltadapprox}
\frac{\delta^{\star}_{\textrm{d,A}}}{1+w^{\star}_{\textrm{d,A}}}\simeq\frac{\delta^{\star}_{\textrm{d,B}}}{1+w^{\star}_{\textrm{d,B}}}\simeq\frac{\delta^{\star}_{\textrm{d,C}}}{1+w^{\star}_{\textrm{d,C}}}.
\end{eqnarray}
Given the model parameters used in this work \cite{Bouali:2019whr}, the EoS parameters at the beginning (deep inside the radiation era) read
\begin{eqnarray}
w^{\star}_{\textrm{d,A}}=-1.027, \:\: w^{\star}_{\textrm{d,B}}=-1.050, \:\:  w^{\star}_{\textrm{d,C}}=-1.320.
\end{eqnarray}
Therefore, the relation of the initial DE perturbation between the different models is roughly
\begin{eqnarray}\label{initialdeltadapproxrel}
&12 \delta^{\star}_{\textrm{d,A}}\simeq\frac{13}{2} \delta^{\star}_{\textrm{d,B}}\simeq\delta^{\star}_{\textrm{d,C}}.
\end{eqnarray}
As can be seen the larger is the deviation  from $-1$ of the  initial EoS parameter, the larger is the initial DE density contrast. This  explains the large initial amplitude for  model C. Similarly, in the case of  model B it can be  observed a weak decay, while in model A, on the contrary, it is almost constant. Despite the large deviation given by these DE models in the early Universe, the amplitudes  are strongly suppressed during the late-radiation dominated epoch and matter domination era, in such a way that different models  predict very similar results at present time, and therefore, no significant deviations should be expected at a future cosmic time.

\subsection{Evolution of the gravitational potential}

\begin{figure*}[h]
 \includegraphics[width=\textwidth]{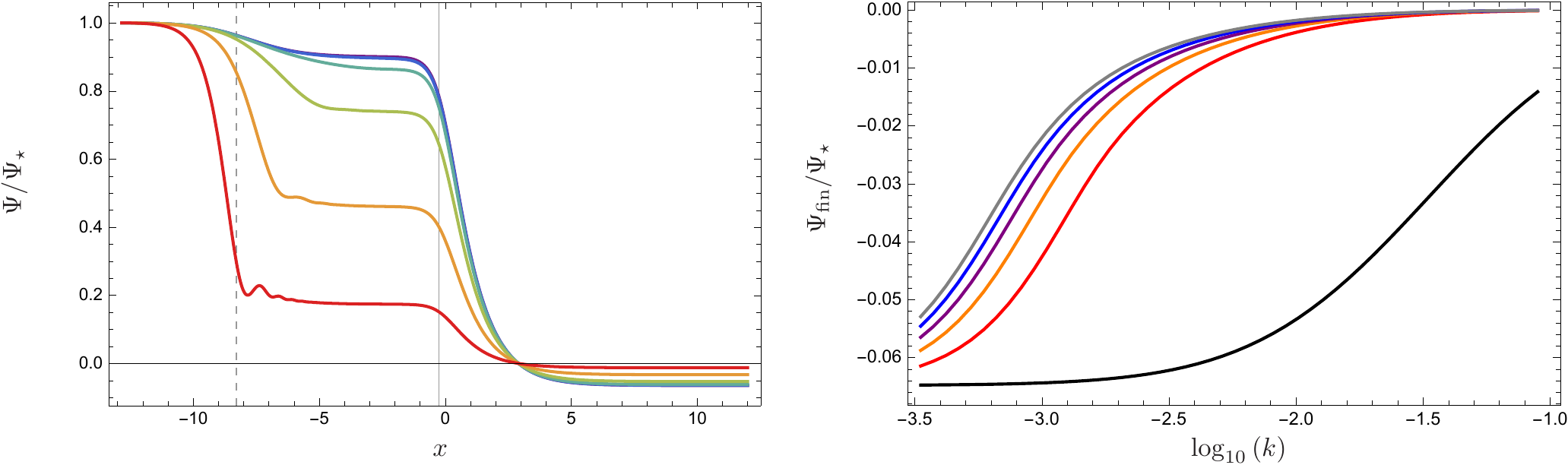}
\caption{The left panel of the  above figure presents the evolution of the gravitational potential divided by its initial value, $\Psi^{\star}$.  These results correspond to model A and choosing a vanishing effective speed of sound $c^{2}_{s\textrm{d}}=0$.  The vertical dashed line corresponds to radiation-matter equality, $x\sim-8.24$, while the solid gray vertical line denotes the matter-DE equality, $x\sim-0.27$. The range and the numerical value of the modes for different colours are the same  as those used in  figure~\ref{deltadall}. The right panel shows the asymptotic value of the gravitational potential in terms of $\log(k)$ for five different values of the effective speed of sound parameter: $c^{2}_{s\textrm{d}}=0$ (black), $c^{2}_{s\textrm{d}}=0.2$ (red), $c^{2}_{s\textrm{d}}=0.4$ (orange), $c^{2}_{s\textrm{d}}=0.6$ (purple),  $c^{2}_{s\textrm{d}}=0.8$ (blue) and $c^{2}_{s\textrm{d}}=1$ (gray).}\label{PsiPsifinal}
\end{figure*}

Aside from DE perturbations analysis, we found some deviations  on the evolution of the gravitational potential.  We remind that in our phantom models the gravitational potential evolves asymptotically to a positive constant, which is not the case of a $\Lambda$CDM or standard DE models, where the gravitational potential evolves towards a vanishing or a negative constant \cite{Albarran:2017kzf}. Since no relevant differences are observed for the different models, we just present the results corresponding to model A. The left panel of 
figure~\ref{PsiPsifinal} shows the  gravitational potential evolution  for a vanishing effective speed of sound parameter.  As can be seen, at a particular scale factor the gravitational potential flips the sign. We can notice that the gravitational potential evolution is almost unaffected by changing $c_{s\textrm{d}}^2$ from the early time till present. However, in the far future some differences merge. 

\

The left panel of figure~\ref{PsiPsifinal} shows the evolution of the gravitational potential, $\Psi$, divided by its initial value, $\Psi_{\star}$ for the six relevant modes previously chosen. As can be seen, the gravitational potential almost vanishes for the largest modes, while it evolves to a positive constant for small modes. We remind once again that the gravitational potential is negative at the beginning of the computations, which confers the attractive nature of gravity. Therefore, a positive sign on the gravitational potentials is understood as a repulsive effect. 

\

The right panel of figure~\ref{PsiPsifinal} shows the asymptotic value of the gravitational potential divide by the initial value, $\Psi_{\star}$. The plot is done to highlight how such a constant is affected by the different values of the modes and $c^{2}_{s\textrm{d}}$ parameter. As can be seen, for large modes the gravitational potential vanishes with independence  of the chosen $c^{2}_{s\textrm{d}}$ parameter, while for the smallest modes such a constant is set to be around $-0.065$. Bear in mind that the initial value of the gravitational potential is negative while asymptotically it approximates to the constant $\Psi_{\textrm{fin}}$, which is a positive value.

\begin{figure*}[h!]
 \includegraphics[width=\textwidth]{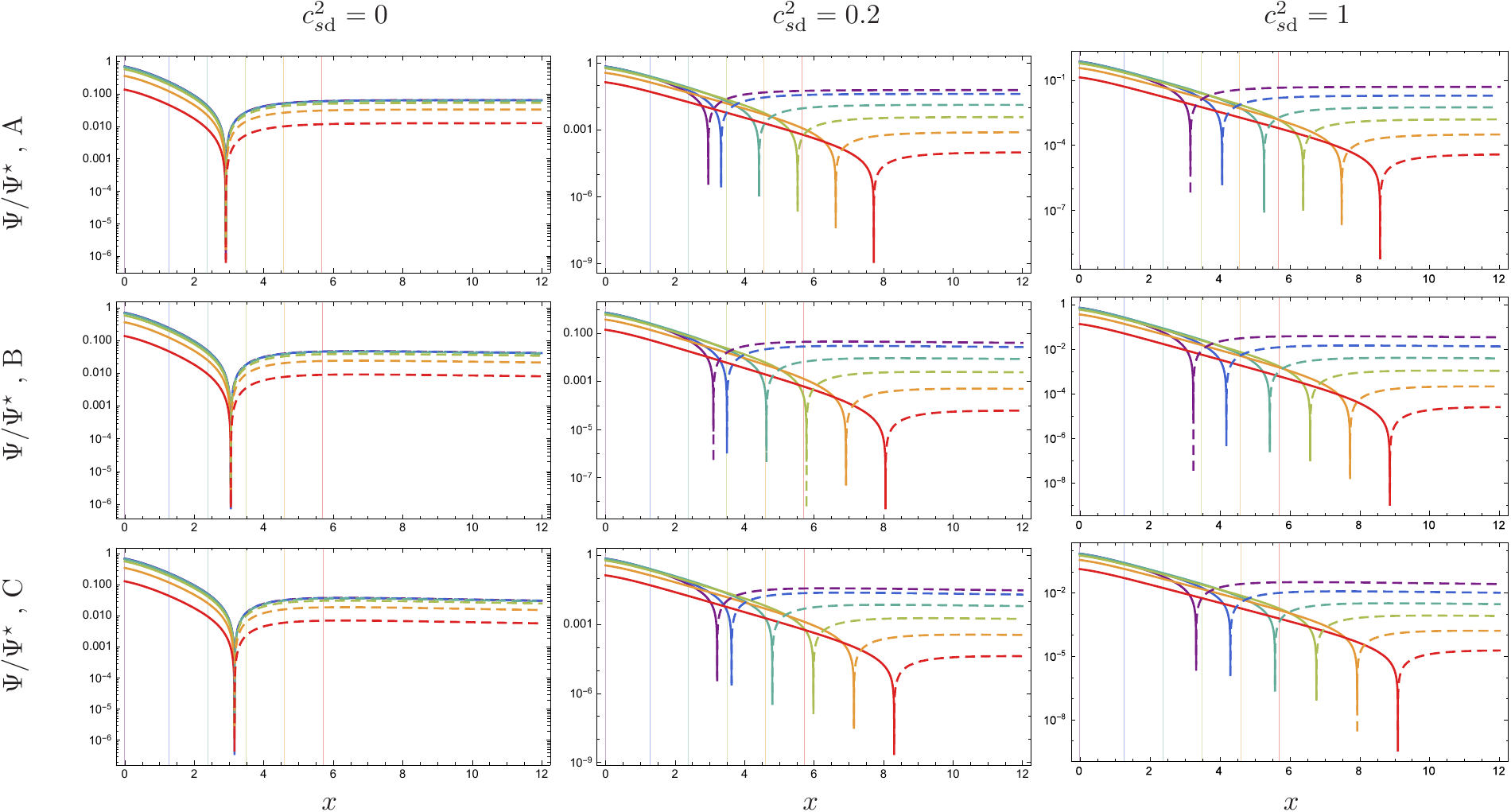}
\caption{This figure presents the evolution of the gravitational potential, $\Psi$,  with respect to its initial value, $\Psi^{\star}$,  in a logarithmic  scale and from the present to the far future. The solid lines represent the positive values while the dashed lines represent negative values. We apply the same criteria as the ones used in figure~\ref{deltadall}  to represent the different modes.  The coloured vertical lines represent the moment of horizon exit for the corresponding mode.}\label{deltapsiall}
\end{figure*}

\

We find interesting to focus on the evolution of the gravitational potential in the far future, mainly, where it flips its sign. For instance, figure \ref{deltapsiall} shows the evolution of the gravitational potential from the present time till the distant future. As can be seen, for a vanishing effective speed of sound parameter the gravitational potential flip of sign occurs, for all the modes, at the same time ($x\sim3$). In addition, such flip occurs before some of the modes have exited the horizon. This is not the case of a  non-vanishing $c_{s\textrm{d}}^2$ parameter (second and third column of figure~\ref{deltapsiall}, for $c_{s\textrm{d}}^{2}=0.2$ and $c_{s\textrm{d}}^{2}=1$, respectively). As can be seen, the smallest modes switch the sign earlier than the largest modes do, however, all the relevant modes have exited the horizon when those flips occur. In addition, we found that that the more abrupt is the cosmic event induced by the model, the sooner occurs the sign flip. This difference is more pronounced the larger are the $k$ modes and $c_{s\textrm{d}}^{2}$ values. 

\

With the aim to better understand the asymptotic evolution of the metric perturbation, we solve the second order differential equation for the gravitational potential.  By incorporating the decomposition  of the pressure (see \cite{Bean:2003fb,Valiviita:2008iv,Ballesteros:2010ks,Arjona:2020yum} for detailed calculations on decomposing the DE pressure in its adiabatic and non-adiabatic contributions) in the perturbation equation of the gravitational potential\footnote{ see, for example, Eq.(3.16) in \cite{Albarran:2016mdu}} we get
\begin{equation}
\begin{split}
\Psi_{xx} + \frac{1}{2}&\left[5-3w+6c^{2}_{a\textrm{d}}\right]\Psi_{x}+\left[3\left(c^{2}_{a\textrm{d}}-w\right)+\frac{c^{2}_{s\textrm{d}}k^2}{\mathcal{H}^2}\right]\Psi=0
\end{split}\label{gravdif1}
\end{equation}
Let us consider a constant EoS parameter where\footnote{The differential  equations for models B and C are not the same. However, after solving  those cases by numerical analysis, we have not found significant deviations with respect the model A.} $c^{2}_{a\textrm{d}}=w_{\textrm{d}}$. Therefore, in a phantom DE dominated Universe the Eq~(\ref{gravdif1}) can be approximated as\footnote{In the case of the models B and C, this assumption is not correct since the differential equation (\ref{gravdif1}) is different. However, we do not observe significant changes between the numerical results given by the different models. So we focus on model A  since its differential equation becomes analytically solvable.} 
\begin{equation}
\begin{split}
\Psi_{xx} + \frac{1}{2}&\left(5+3w_{\textrm{d}}\right)\Psi_{x}+\left(\frac{c^{2}_{s\textrm{d}}k^2}{\Omega_{\textrm{d}0}k^2_{0}}\right)e^{\left(3w_{\textrm{d}}+1\right)x} \ \Psi=0,
\end{split}\label{gravdif2}
\end{equation}
whose solutions are given by 
\begin{equation}\label{gravdifxlarge1}
\Psi_{\left[c^{2}_{s\textrm{d}}=0\right]}=C_{1}+\frac{C_{2}}{\beta}e^{-\beta x},
\end{equation}
\begin{equation}
\begin{split}
\Psi_{\left[c^{2}_{s\textrm{d}}\neq0\right]}=e^{-\frac{\beta}{2}x}
&\Bigl\{D_{1}J_{\nu}\left[A\left(k\right) e^{-\gamma x}\right]-D_{2}Y_{\nu}\left[A\left(k\right) e^{-\gamma x}\right]\Bigr\},
\end{split}\label{gravdifxlarge2}
\end{equation}
\begin{equation}
\begin{split}
\Psi_{\left[c^{2}_{s\textrm{d}}\neq0\right]}\sim &\frac{D_{1}}{\Gamma\left(\nu+1\right)}\left[\frac{A\left(k\right)}{2}\right]^{\nu}+\frac{D_{2}\Gamma\left(-\nu\right)}{\pi}\left[\frac{A\left(k\right)}{2}\right]^{-\nu}e^{-\beta x}\:\: \textrm{for} \ \ 1\ll x,
\end{split}\label{gravdifxlarge3}
\end{equation}
where $J_{\nu}$ and $Y_{\nu}$  are the first kind Bessel functions  with order $\nu$, $\Gamma$ is the Gamma function, while $C_{1}$, $C_{2}$, $D_{1}$ and $D_{2}$ are integration constants. The remaining parameters are defined as 
\begin{equation}
\beta\equiv\frac{1}{2}\left(5+3w_{\textrm{d}}\right) ,  \qquad \gamma\equiv-\frac{1}{2}\left(1+3w_{\textrm{d}}\right), \qquad 
\nu\equiv-\frac{\beta}{2\gamma}, \qquad A\left(k\right)\equiv\frac{1}{\gamma}\sqrt{\frac{c^{2}_{s\textrm{d}}}{\Omega_{\textrm{d}0}}}\frac{k}{k_{0}}.
\end{equation}
Since $\Psi$ is linear, a particular solution multiplied by a constant factor is still a solution. Therefore, the total result can be written as
\begin{eqnarray}\label{totalsol}
\Psi_{\textrm{tot}}\left(x\right)=\Psi\left(x\right)F\left(k,c_{s\textrm{d}}\right),
\end{eqnarray}
where $F\left(k,c_{s\textrm{d}}\right)$ can be fixed (with an appropriate choice for $C_{1}$, and $D_{1}$) by analysing the asymptotic behaviour of the gravitational potential shown on the right panel plot of figure~\ref{PsiPsifinal}. 

\

As can be seen, the asymptotic behaviour for large scale factors, given in\footnote{Notice that the coefficient $\gamma$ is positive, therefore, the argument of Bessel function vanishes when $x\rightarrow\infty$. We have obtained the expression for small arguments  making use of   (9.1.7) and (9.1.9) of reference \cite{Abramowitz})} Eq~(\ref{gravdifxlarge3}), coincides with the solution for a vanishing $c_{s\textrm{d}}^{2}$, Eq~(\ref{gravdifxlarge1}). However, instead of having just the constants terms $C_{1}$, and $C_{2}$, the solutions for non-vanishing $c_{s\textrm{d}}^{2}\neq0$ parameter keep some information of the modes through the function $A(k)$ and modulated by the constants $D_{1}$, and $D_{2}$. Note that the dominant solution for $x\rightarrow\infty$  is constant as long as the coefficient $\beta$ is positive, i.e. $-5/3<w_{\textrm{d}}$ which is indeed our case. 

\

Finally, in order to obtain the point where the gravitational potential flip of sign occurs, we just solve $\Psi=0$ for the couple of equations (\ref{gravdifxlarge1}) and (\ref{gravdifxlarge2}). Therefore, we get
\begin{equation}
x_{\textrm{crit}}=-\frac{1}{\beta}\ln\left[-\alpha_{1}\beta\right],\label{xflipsolcrit}
\end{equation}
\begin{equation}
x_{\textrm{flip}}=-\frac{1}{\beta}\ln\left[\alpha_{2}\sin\left(\pi\nu\right)\right]+\frac{1}{\gamma}\ln\left[\frac{1}{2}\sqrt{\frac{c^{2}_{s\textrm{d}}}{\Omega_{\textrm{d}0}}}\right]
+\frac{1}{\gamma}\ln\left[\frac{k}{k_{0}}\right].
\label{xflipsolflip}
\end{equation}
where we have defined a proportionality between the integration constants, i.e. $C_{1}/C_{2}\equiv\alpha_{1}$ and $D_{1}/D_{2}\equiv\alpha_{2}$. Given that $C_{1},D_{1}<0$ and $0<C_{2},D_{2}$, $\alpha_{1}$ and $\alpha_{2}$ are negative constants.

\ 

On the one  hand, $x_{\textrm{crit}}$ is the lower value for which the gravitational potential can switch its sign and corresponds to a vanishing effective speed of sound parameter. Bear in mind that the  differential equation (\ref{gravdif2}) remains invariant by choosing different $k$ and $c^{2}_{s\textrm{d}}$ as long as the product $k^{2}c^{2}_{s\textrm{d}}$ is fixed. Therefore, the solution  for the  limit $k\rightarrow0$ corresponds  to the solution for a vanishing $c^{2}_{s\textrm{d}}$. On the other hand, for non-vanishing values of the product $k^{2}c^{2}_{s\textrm{d}}$, the  moment at which the gravitational potential flips its sign is given by  (\ref{xflipsolflip}) (which is valid as long as $x_{\textrm{crit}}<x_{\textrm{flip}}$). 

\

Therefore, we could define a second surface whose size is the distance where the gravitational potential becomes positive. In addition, we notice that such a second surface changes with time as fast as the true horizon does, i.e. $\ln{\left(k/k_{0}\right)}\sim\gamma x$. So this could be understood as two horizons, the true one; i.e. that enclose the observable Universe, and a second one; i.e. where the gravitational potential becomes positive, keeping the relative distance as constant.

\

\begin{figure*}[h!]
 \centering 
 \includegraphics[width=10.5cm]{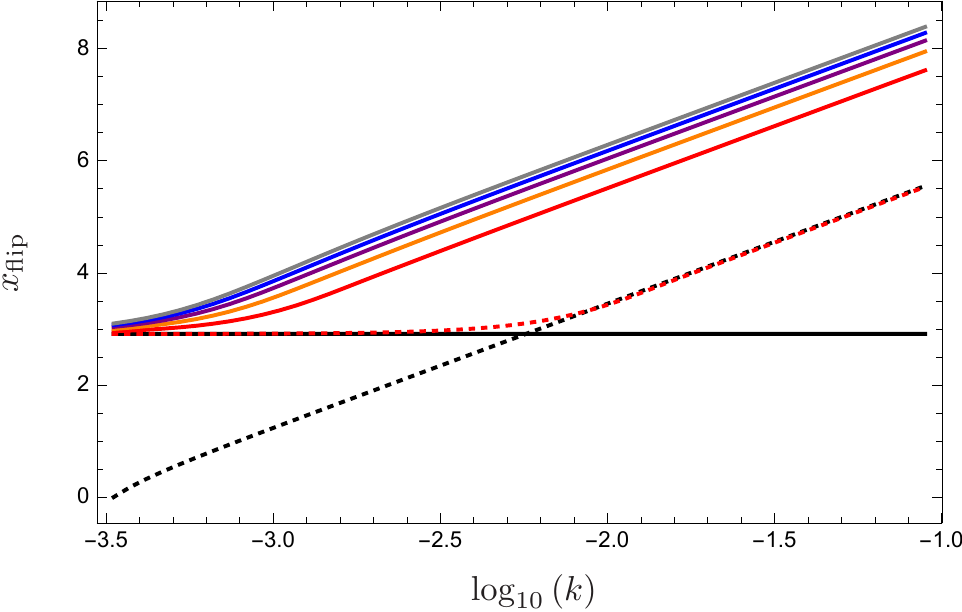}
\caption{This figure shows the moment where the gravitational potential switches its sign, $x_{\textrm{flip}}$, in terms of $\log_{10}\left(k\right)$. Each curve corresponds to a given value of the speed of sound parameter; $c^{2}_{s\textrm{d}}=0$ (solid black), $c^{2}_{s\textrm{d}}=0.2$ (red), $c^{2}_{s\textrm{d}}=0.4$ (orange), $c^{2}_{s\textrm{d}}=0.6$ (purple),$c^{2}_{s\textrm{d}}=0.8$ (blue) and  $c^{2}_{s\textrm{d}}=1$ (gray). The red dashed curve corresponds to $c^{2}_{s\textrm{d}}=2.09\times 10^{-3}$ while the black-dotted line represent the value of $x$ where modes exit the horizon.}\label{xflip}
\end{figure*}

Figure~\ref{xflip} shows a plot of $x_{\textrm{flip}}$ vs $\log_{10}\left(k\right)$. Since no relevant differences are found between models, we again only present the result given for the model A. We should keep in mind that the differential equation that lead to the analytical solution will be different if instead we choose model B or C. However, in practical terms all the models give similar numerical solutions.

\

As it is shown in Figure~\ref{xflip}, for a vanishing $c^{2}_{s\textrm{d}}$ parameter (solid black line) the flip of sign occurs at the same time for all the modes. The plots given by non-vanishing $c^{2}_{s\textrm{d}}$ parameters (coloured solid lines) becomes parallel, at  large modes, with respect to the horizon exiting line (black dotted line). This means that there is an upper bound on $c^{2}_{s\textrm{d}}$ that will ensure sign-flipped modes inside the horizon, while values larger than such an upper bound would stand beyond the observable Universe. We have estimated such an upper bound roughly to be the order of  $c^{2}_{s\textrm{d}}\lesssim2\times 10^{-3}$ (represented by the red dotted curve).

%

%
%
%
%
\section{Conclusions}\label{conclusions_Speed}

In this work, we have analysed the cosmological perturbations of three genuine phantom DE models with a varying effective speed of sound parameter. These models, named  in the present work as model A,  B and  C,  induce a particular future event known as BR, LR and LSBR, respectively. In these future events the Universe reach a scenario where all the bound structures are ripped apart.  We have addressed the computation of the linear cosmological perturbations following the method of decomposing the DE pressure perturbation in its adiabatic and non adiabatic contributions \cite{Bean:2003fb,Valiviita:2008iv}, which leaves a dynamical set of equations free of instabilities. In this way, the effective speed of sound parameter of DE, $c^{2}_{s\textrm{d}}$, is regarded as a free parameter. 

\

We have considered a Universe filled with radiation, matter and DE, where the latter is described by the aforementioned models. We have computed the perturbations  since the radiation dominated epoch, ($a_{\textrm{ini}}\sim2.6\times10^{-6}$), till a far future ($a_{\textrm{fin}}\sim1.6\times10^{5}$), where DE completely dominates. On the one hand, the  model parameters where fixed by using the background observational constraints obtained in \cite{Bouali:2019whr}. On the other hand, the physical values as the initial conditions for single field inflation, giving rise to the spectral amplitude and spectral index were fixed using Planck data \cite{wikiesa}. Then, we obtain the predicted current matter power spectrum and the evolution of $f\sigma_{8}$ growth rate. Finally, we study the effect of changing $c^{2}_{s\textrm{d}}$ from $0$ to $1$. 

\

We find that different values of the $c^{2}_{s\textrm{d}}$ parameter does not affect significantly the matter perturbations. Consequently, the matter power spectrum and $f\sigma_{8}$ evolution do not show any relevant footprint. In fact, the relative deviations with respect to $c^{2}_{s\textrm{d}}=1$ are, in the best case, up to $10^{-2}$ in the matter power spectrum (for small modes) and $10^{-3}$ in $f\sigma_{8}$ (for vanishing values of $c^{2}_{s\textrm{d}}$). Given that the upcoming Euclid data is expected to measure the primordial matter power spectrum with an accuracy of one percent \cite{Laureijs:2011gra}, we expect that the effects predicted in this work  will be difficult, but  not impossible, to detect. A combined data set of large scale structure with Planck could increase the accuracy  just enough to distinguish the footprints predicted at small modes.  On the one hand, the accuracy on the growth rate might not be enough to measure the very small deviations predicted by our results (see figure~2.5 in reference \cite{Laureijs:2011gra}). On the other hand, the associated deviations into the gravitational potential, $\Psi$, due to the different values of the $c^{2}_{s\textrm{d}}$ parameter become important at large scales, reaching the largest deviation at asymptotic values (see right panel in figure~ \ref{PsiPsifinal}). Nevertheless, we expect that at present time those deviations in $\Psi$ are very small (see figure~\ref{deltapsiall} at $x=0$) to induce a significant change when computing the cross correlation between CMB and the large scale structure in pursuit of footprints traced by $c^{2}_{s\textrm{d}}$. The integrated Sachs-Wolf effect has its footprints mainly connected with the gravitational potential behaviour, which is more sensitive to the chosen models, due to its  dependence on the adiabatic speed of sound, $c^{2}_{a\textrm{d}}$, rather than the value of $c^{2}_{s\textrm{d}}$. So we expect that the effects of different values of $c^{2}_{s\textrm{d}}$ will leave no significant footprints detectable on the integrated Sachs-Wolf effect.  However, we have found interesting footprints  in the DE density contrast when changing $c^{2}_{s\textrm{d}}$. Those changes are  amplified when $c^{2}_{s\textrm{d}}$ is set very small. Despite the different three models are almost indistinguishable at present, there are significant deviations in the  early Universe, which strongly depend on the initial EoS parameter of DE  due to the adiabatic conditions imposed at the beginning (see   (\ref{initialdeltadapprox}) and (\ref{initialdeltadapproxrel})).

\

We conclude that the possibility of a vanishing speed of sound parameter does not seem to be  favoured by two reasons: (i) the DE density contrast grows too much during the matter dominated epoch, and this would lead to a DE clustering, something that has not been detected so far, (ii) the gravitational potential sign flip occurs at the same time for all the modes, such unexpected and sudden event does not seem physical. This is not the case of a non-vanishing effective speed of sound parameter, where the Bardeen potential becomes   progressively negative from very large distances to smaller ones. Such distances decrease with time as fast as the horizon does. Therefore, there is a particular value of $c^{2}_{s\textrm{d}}$ where the gravitational potential switches the sign precisely at the horizon. We have found that this value is close to $c^{2}_{s\textrm{d}}\sim 2\times 10^{-3}$. For such a small value the DE clustering could be large enough to become detectable, a fact that has not been observed yet. Therefore, it could play an important role favouring or ruling out different DE models and $c^{2}_{s\textrm{d}}$ values. We hope the upcoming Euclid mission will provide a refined data on the $c^{2}_{s\textrm{d}}$ parameter likewise other important cosmological parameters. We have found that the standard deviation Euclid will present on the DE speed of sound parameter is large for values close to one, $1\ll\sigma(c^{2}_{s\textrm{d}})/c^{2}_{s\textrm{d}}$ when $c^{2}_{s\textrm{d}}\sim1$, while it gets a reasonable accuracy close to vanishing values, $\sigma(c^{2}_{s\textrm{d}})/c^{2}_{s\textrm{d}}\sim0.11$ when $c^{2}_{s\textrm{d}}=10^{-6}$ (see page 143 on reference \cite{Amendola:2016saw}). Therefore, if the measured speed of sound is non-zero but close to vanishing values, then it could become a suitable indicator to discriminate or favour different DE models, aside from the potential use to constrain several DE models.

\

Despite  the fact that the DE perturbations have not been observed so far, we strongly believe  that they hide revealing footprints that could allow us to distinguish different DE models if ever detected. We hope that upcoming missions such as Euclid will provide crucial  information about the dark sector of the Universe, granting us a useful tool to favouring or discriminating among DE models as those addressed in the present work.

%
%
%
%
%

\section*{Acknowledgments}
The research of M. B.-L. is supported by the Basque Foundation of Science Ikerbasque. 
She also would like to acknowledge the support from the Basque
government Grant No. IT956- 16 (Spain) and from the Grant PID2020-114035GB-100 funded by
MCIN/AEI/10.13039/501100011033 and by “ERDF A way of making Europe”. She is as well thankful to CMA-UBI (Portugal) for kind hospitality while doing part of this work.
IA and JM are supported by the grant UIDB/MAT/00212/2020.
%
%
%
%
%
\appendix
\section{Decomposition of a non-adiabatic pressure}\label{pressure_decompose}
We first consider a gauge transformation from the rest frame to the Newtonian gauge. Therefore, the physical quantities in both gauges are related as 
\begin{equation}
\begin{split}
\delta p_{\ell}&=\left.\delta p_{\ell}\right|_{\textrm{r.f.}}- p'_{\ell}\delta\eta , \quad
\delta \rho_{\ell}=\left.\delta \rho_{\ell}\right|_{\textrm{r.f.}}- \rho'_{\ell}\delta\eta \\ 
\left(v_{\ell}+B\right)&=\left.\left(v_{\ell}+B\right)\right|_{\textrm{r.f.}}+\delta\eta,
\end{split}\label{transgauge}
\end{equation}
where  subscript  r.f denotes  rest frame and no subscript refers to Newtonian gauge. In the rest frame, $v_{\ell\textrm{r.f.}}=0$ and $B_{\textrm{r.f.}}=0$, while in the Newtonian gauge it is set $B=0$. Therefore, this implies $v_{\ell}=\delta\eta$. On the other hand, we consider that the total pressure perturbation is given by its adiabatic and non-adiabatic contributions, 
\begin{equation}\label{adplusnonad}
\delta p_{\ell}=\delta p_{\ell\textrm{ad}}+\delta p_{\ell\textrm{nad}},
\end{equation}
where by  definition, $\delta p_{\ell\textrm{ad}}=c^{2}_{a\ell}\delta\rho_{\ell}$ is the adiabatic part.  We compute the gauge difference on both sides of the equality (\ref{adplusnonad}) using Eq.(~\ref{transgauge}) together with the definitions for the speed of sounds
\begin{eqnarray}\label{newvar0}
	c_{sA}^{2}=\left.\frac{\delta p_A}{\delta\rho_{A}}\right\vert_\textrm{r.f.}
	\,,
	\qquad 
	c_{aA}^2=\frac{p_A'}{\rho_A'}
	\,,
\end{eqnarray}
 First, we deduce that  the non adiabatic part is gauge invariant, i.e. $\delta p_{\ell\textrm{nad}}=\delta p_{\ell\textrm{nad}}\vert_{\textrm{r.f.}}$. Therefore, we get
\begin{equation}
\begin{split}
\delta p_{\ell\textrm{nad}}&=\left(c_{s\ell}^{2}-c_{a\ell}^{2}\right)\left.\delta \rho_{\ell}\right|_{\textrm{r.f.}}, \\
\delta p_{\ell}&=c_{s\ell}^{2}\delta \rho_{\ell}+\left(c_{s\ell}^{2}-c_{a\ell}^{2}\right)\rho'_{\ell}\delta\eta.
\end{split}\label{deltapnonadiab}
\end{equation}
Finally, making use of the conservation equation, $\rho'_{\ell}=-3\mathcal{H}\left(1+w_{\ell}\right)\rho_{\ell}$, the non-adiabatic contribution and total pressure perturbation can be  written as 
\begin{equation}
\begin{split}
\delta p_{\ell\textrm{nad}}&=\left(c_{s\ell}^{2}-c_{a\ell}^{2}\right)\left[\delta \rho_{\ell}-3\mathcal{H}\left(1+w_{\ell}\right)\rho_{\ell}v_{\ell}\right], \\
\delta p_{\ell}&=c_{s\ell}^{2}\delta\rho_{\ell}+\left(c_{a\ell}^{2}-c_{s\ell}^{2}\right)3\mathcal{H}\left(1+w_{\ell}\right)\rho_{\ell}v_{\ell}.
\end{split}\label{difpone}
\end{equation}
%
%
%
%
%


\begin{thebibliography}{99}
%
%
%
\bibitem{Riess:1998cb}
 A.~G.~Riess {\it et al.} [Supernova Search Team Collaboration],
 Astron.\ J.\ {\bf 116} (1998) 1009
 [\href{https://arxiv.org/abs/astro-ph/9805201}{astro-ph/9805201}].
 
\bibitem{Perlmutter:1998np}
 S.~Perlmutter {\it et al.} [Supernova Cosmology Project Collaboration],
 Astrophys.\ J.\ {\bf 517} (1999) 565
 [\href{https://arxiv.org/abs/astro-ph/9812133}{astro-ph/9812133}].
 
  \bibitem{AmendolaTsujikawa}
 L. Amendola and S. Tsujikawa,
 \textit{Dark Energy: Theory and Observations}. First edition (Cambridge University Press, Cambridge, 2010). 
 



 
\bibitem{Caldwell:2003vq}
 R.~R.~Caldwell, M.~Kamionkowski and N.~N.~Weinberg,
 Phys.\ Rev.\ Lett.\ {\bf 91} (2003) 071301
 [\href{https://arxiv.org/abs/astro-ph/0302506}{astro-ph/0302506}].
 
  
\bibitem{Caldwell:1999ew}
 R.~R.~Caldwell,
 Phys.\ Lett.\ B {\bf 545} (2002) 23
 [\href{http://arxiv.org/abs/astro-ph/9908168}{astro-ph/9908168}].
 

  
\bibitem{Dabrowski:2003jm}
 M.~P.~D\c{a}browski, T.~Stachowiak, and M.~Szyd{\l }owski,
 Phys.\ Rev.\ D {\bf 68} (2003) 103519
 [\href{https://arxiv.org/abs/hep-th/0307128}{hep-th/0307128}].
 
  
\bibitem{Stefancic:2003rc}
  H.~\^{S}tefan\^{c}i\'{c},
  Phys.\ Lett.\ B {\bf 586} (2004) 5
 [\href{ https://arxiv.org/abs/astro-ph/0310904}{arXiv:0310904[astro-ph.CO]}].
 


 
 
\bibitem{DiValentino:2016hlg}
E.~Di Valentino, A.~Melchiorri and J.~Silk,
Phys. Lett. B \textbf{761} (2016), 242-246
 [\href{https://arxiv.org/abs/1606.00634}{arXiv:1606.00634 [astro-ph.CO]}].
 
\bibitem{Vagnozzi:2018jhn}
S.~Vagnozzi, S.~Dhawan, M.~Gerbino, K.~Freese, A.~Goobar and O.~Mena,
Phys. Rev. D \textbf{98} (2018) no.8, 083501
  [\href{https://arxiv.org/abs/1801.08553}{arXiv:1801.08553 [astro-ph.CO]}].
  
  
\bibitem{Vagnozzi:2019ezj}
S.~Vagnozzi,
Phys. Rev. D \textbf{102} (2020) no.2, 023518
 [\href{https://arxiv.org/abs/1907.07569}{arXiv:1907.07569 [astro-ph.CO]}].
  
\bibitem{Alestas:2020mvb}
G.~Alestas, L.~Kazantzidis and L.~Perivolaropoulos,
Phys. Rev. D \textbf{101} (2020) no.12, 123516
[\href{https://arxiv.org/abs/2004.08363}{arXiv:2004.08363 [astro-ph.CO]}].
 
 
\bibitem{DiValentino:2020vnx}
E.~Di Valentino,
Mon. Not. Roy. Astron. Soc. \textbf{502} (2021) no.2, 2065-2073
[\href{https://arxiv.org/abs/2011.00246}{arXiv:2011.00246 [astro-ph.CO]}].

\bibitem{DiValentino:2020naf}
E.~Di Valentino, A.~Mukherjee and A.~A.~Sen,
Entropy \textbf{23} (2021) no.4, 404
[\href{https://arxiv.org/abs/2005.12587}{arXiv:2005.12587 [astro-ph.CO]}].

\bibitem{Yang:2021hxg}
W.~Yang, S.~Pan, E.~Di Valentino, O.~Mena and A.~Melchiorri,
[\href{https://arxiv.org/abs/2101.03129}{arXiv:2101.03129 [astro-ph.CO]}].

\bibitem{Yang:2021egn}
W.~Yang, E.~Di Valentino, S.~Pan and O.~Mena,
Phys. Dark Univ. \textbf{31} (2021), 100762
[\href{https://arxiv.org/abs/2007.02927}{arXiv:2007.02927 [astro-ph.CO]}].

\bibitem{DiValentino:2021izs}
E.~Di Valentino, O.~Mena, S.~Pan, L.~Visinelli, W.~Yang, A.~Melchiorri, D.~F.~Mota, A.~G.~Riess and J.~Silk,
 [\href{https://arxiv.org/abs/2103.01183}{arXiv:2103.01183 [astro-ph.CO]}].
 

 
  
\bibitem{Aghanim:2018eyx}
  N.~Aghanim {\it et al.} [Planck Collaboration],
 [\href{https://arxiv.org/abs/1807.06209}{arXiv:1807.06209 [astro-ph.CO]}].
 
 
\bibitem{Starobinsky:1999yw}
 A.~A.~Starobinsky,
 Grav.\ Cosmol.\ {\bf 6} (2000) 157
 [\href{http://arxiv.org/abs/astro-ph/9912054}{astro-ph/9912054}].
 
 
\bibitem{Carroll:2003st}
 S.~M.~Carroll, M.~Hoffman and M.~Trodden,
 Phys.\ Rev.\ D {\bf 68} (2003) 023509
 [\href{http://arxiv.org/abs/astro-ph/0301273}{astro-ph/0301273}].
 

 
 



 
\bibitem{Chimento:2003qy}
 L.~P.~Chimento and R.~Lazkoz,
 Phys.\ Rev.\ Lett.\ {\bf 91} (2003) 211301
 [\href{http://arxiv.org/abs/gr-qc/0307111}{gr-qc/0307111}].

\bibitem{GonzalezDiaz:2003rf}
 P.~F.~Gonz\'{a}lez-D\'{i}az,
 Phys.\ Lett.\ B {\bf 586} (2004) 1
 [\href{http://arxiv.org/abs/astro-ph/0312579}{astro-ph/0312579}].

\bibitem{GonzalezDiaz:2004vq}
 P.~F.~Gonz\'{a}lez-D\'{i}az,
 Phys.\ Rev.\ D {\bf 69} (2004) 063522
 [\href{https://arxiv.org/abs/hep-th/0401082}{hep-th/0401082}].
 
 \bibitem{Ruzmaikina}
T. Ruzmaikina and A. A. Ruzmaikin, Sov. Phys. JETP {\bf 30} (1970) 372.
 
 
\bibitem{Bouhmadi-Lopez:2013nma}
 M.~Bouhmadi-L\'{o}pez, P.~Chen, and Y.~W.~Liu,
 Eur.\ Phys.\ J.\ C {\bf 73} (2013) 2546
 [\href{https://arxiv.org/abs/1302.6249}{arXiv:1302.6249 [gr-qc]}].

 
\bibitem{Nojiri:2005sx}
 S.~'i.~Nojiri, S.~D.~Odintsov and S.~Tsujikawa,
 Phys.\ Rev.\ D {\bf 71} (2005) 063004
 [\href{https://arxiv.org/abs/hep-th/0501025}{hep-th/0501025}].
 
\bibitem{Nojiri:2005sr}
 S.~'i.~Nojiri and S.~D.~Odintsov,
 Phys.\ Rev.\ D {\bf 72} (2005) 023003
 [\href{https://arxiv.org/abs/hep-th/0505215}{hep-th/0505215}].
 
\bibitem{Stefancic:2004kb}
 H.~\v{S}tefan\v{c}i{\'c},
 Phys.\ Rev.\ D {\bf 71} (2005) 084024
 [\href{https://arxiv.org/abs/astro-ph/0411630}{astro-ph/0411630}].
 
\bibitem{BouhmadiLopez:2005gk}
 M.~Bouhmadi-L{\'o}pez,
 Nucl.\ Phys.\ B {\bf 797} (2008) 78
 [\href{https://arxiv.org/abs/astro-ph/0512124}{astro-ph/0512124}].
 
\bibitem{Frampton:2011sp}
 P.~H.~Frampton, K.~J.~Ludwick and R.~J.~Scherrer,
 Phys.\ Rev.\ D {\bf 84} (2011) 063003
 [\href{https://arxiv.org/abs/1106.4996}{arXiv:1106.4996 [astro-ph.CO]}].
 
\bibitem{Brevik:2011mm}
 I.~Brevik, E.~Elizalde, S.~'i.~Nojiri, and S.~D.~Odintsov,
 Phys.\ Rev.\ D {\bf 84} (2011) 103508
 [\href{https://arxiv.org/abs/1107.4642}{arXiv:1107.4642 [hep-th]}].
 
\bibitem{Bouhmadi-Lopez:2014cca}
 M.~Bouhmadi-L\'opez, A.~Errahmani, P.~Mart\'{i}n-Moruno, T.~Ouali, and Y.~Tavakoli,
 Int.\ J.\ Mod.\ Phys.\ D {\bf 24} (2015) no.10, 1550078
 [\href{https://arxiv.org/abs/1407.2446}{arXiv:1407.2446 [gr-qc]}].
 
\bibitem{Bouhmadi-Lopez:2018lly}
  M.~Bouhmadi-L\'{o}pez, D.~Brizuela and I.~Garay,
  JCAP {\bf 1809} (2018) no.09,  031
  [\href{https://arxiv.org/abs/1802.05164}{ [arXiv:1802.05164 [gr-qc]]}].
 
\bibitem{Morais:2016bev}
 J.~Morais, M.~Bouhmadi-L\'opez, K.~Sravan Kumar, J.~Marto and Y.~Tavakoli,
 [\href{http://arxiv.org/abs/1608.01679}{arXiv:1608.01679 [gr-qc]}].
 
  
\bibitem{Laureijs:2011gra}
  R.~Laureijs {\it et al.} [EUCLID Collaboration],
   [\href{https://arxiv.org/abs/1110.3193}{arXiv:1110.3193 [astro-ph.CO]}].
  
\bibitem{Amendola:2016saw}
  L.~Amendola {\it et al.},
  Living Rev.\ Rel.\  {\bf 21} (2018) no.1,  2
   [\href{https://arxiv.org/abs/1606.00180}{arXiv:1606.00180 [astro-ph.CO]}].
   
  
\bibitem{Bean:2003fb}
 R.~Bean and O.~Dor\'{e},
 Phys.\ Rev.\ D {\bf 69} (2004) 083503
 [\href{https://arxiv.org/abs/astro-ph/0307100}{astro-ph/0307100}].
 
\bibitem{Valiviita:2008iv}
 J.~V\"{a}liviita, E.~Majerotto and R.~Maartens,
 JCAP {\bf 0807} (2008) 020
 [\href{https://arxiv.org/abs/0804.0232}{arXiv:0804.0232 [astro-ph]}].
 
\bibitem{DeDeo:2003te}
S.~DeDeo, R.~R.~Caldwell and P.~J.~Steinhardt,
Phys. Rev. D \textbf{67} (2003), 103509
[erratum: Phys. Rev. D \textbf{69} (2004), 129902]
 [\href{https://arxiv.org/abs/astro-ph/0301284}{arXiv:astro-ph/0301284 [astro-ph]}].
 
\bibitem{ul:2014dxa}
  H.~A.~Rizwan ul and S.~Unnikrishnan,
  J.\ Phys.\ Conf.\ Ser.\  {\bf 484} (2014) 012048
 [\href{https://arxiv.org/abs/1407.4079}{arXiv:1407.4079 [astro-ph.CO]}].
 
\bibitem{Perkovic:2020eju}
D.~Perkovic and H.~Stefancic,
 [\href{https://arxiv.org/abs/2009.08680}{arXiv:2009.08680 [astro-ph.CO]}].
 
\bibitem{Frusciante:2019xia}
N.~Frusciante and L.~Perenon,
Phys. Rept. \textbf{857} (2020), 1-63 
  [\href{https://arxiv.org/abs/1907.03150}{arXiv:1907.03150 [astro-ph.CO]}].

  
\bibitem{Nesseris:2015fqa}
  S.~Nesseris and D.~Sapone,
  Phys.\ Rev.\ D {\bf 92} (2015) no.2,  023013
 [\href{https://arxiv.org/abs/1505.06601}{1505.06601 [astro-ph.CO]}].
 
\bibitem{Pietrobon:2008js}
  D.~Pietrobon, A.~Balbi, M.~Bruni and C.~Quercellini,
  Phys.\ Rev.\ D {\bf 78} (2008) 083510
 [\href{https://arxiv.org/abs/0807.5077}{arXiv:0807.5077 [astro-ph]}].
 
\bibitem{Ballesteros:2010ks}
  G.~Ballesteros and J.~Lesgourgues,
  JCAP {\bf 1010} (2010) 014
 [\href{https://arxiv.org/abs/1004.5509}{arXiv:1004.5509 [astro-ph.CO]}].
 
\bibitem{Linton:2017ged}
  M.~S.~Linton, A.~Pourtsidou, R.~Crittenden and R.~Maartens,
  JCAP {\bf 1804} (2018) no.04,  043
 [\href{https://arxiv.org/abs/1711.05196}{arXiv:1711.05196 [astro-ph.CO]}].
 
\bibitem{Koivisto:2005mm}
  T.~Koivisto and D.~F.~Mota,
  Phys.\ Rev.\ D {\bf 73} (2006) 083502
   [\href{https://arxiv.org/abs/astro-ph/0512135}{astro-ph/0512135}].
   
\bibitem{Arora:2020lsr}
S.~Arora, X.~h.~Meng, S.~K.~J.~Pacif and P.~K.~Sahoo,
Class. Quant. Grav. \textbf{37}, no.20, 205022 (2020)
[\href{https://arxiv.org/abs/2007.07717}{arXiv:2007.07717 [gr-qc]}].

\bibitem{Velten:2017mtr}
H.~Velten and R.~Fazolo,
Phys. Rev. D \textbf{96} (2017) no.8, 083502
[\href{https://arxiv.org/abs/1707.03224}{arXiv:1707.03224 [astro-ph.CO]}].
 
\bibitem{Arjona:2020yum}
R.~Arjona, J.~Garc\'\i{}a-Bellido and S.~Nesseris,
[\href{https://arxiv.org/abs/2006.01762}{arXiv:2006.01762 [astro-ph.CO]}].


\bibitem{Arjona:2020kco}
R.~Arjona and S.~Nesseris,
   [\href{https://arxiv.org/abs/2001.11420}{arXiv:2001.11420 [astro-ph.CO]}].
   
\bibitem{Niedermann:2019olb}
F.~Niedermann and M.~S.~Sloth,
[\href{https://arxiv.org/abs/1910.10739}{arXiv:1910.10739 [astro-ph.CO]}].

\bibitem{Chudaykin:2020acu}
A.~Chudaykin, D.~Gorbunov and N.~Nedelko,
JCAP \textbf{08}, 013 (2020)
[\href{https://arxiv.org/abs/2004.13046}{arXiv:2004.13046 [astro-ph.CO]}].



  
\bibitem{Bhattacharyya:2019lvg}
A.~Bhattacharyya and S.~Pal,
[\href{https://arxiv.org/abs/1907.10946}{arXiv:1907.10946 [astro-ph.CO]}].

 
\bibitem{Verde:2016wmz}
L.~Verde, E.~Bellini, C.~Pigozzo, A.~F.~Heavens and R.~Jimenez,
JCAP \textbf{04} (2017), 023
[\href{https://arxiv.org/abs/1611.00376}{arXiv:1611.00376 [astro-ph.CO]}]. 
 
\bibitem{dePutter:2010vy}
  R.~de Putter, D.~Huterer and E.~V.~Linder,
  Phys.\ Rev.\ D {\bf 81} (2010) 103513
   [\href{https://arxiv.org/abs/1002.1311}{arXiv:1002.1311 [astro-ph.CO]}].
   
\bibitem{Hu:2004yd}
W.~Hu and R.~Scranton,
Phys. Rev. D \textbf{70} (2004), 123002
 [\href{https://arxiv.org/abs/astro-ph/0408456}{[arXiv:astro-ph/0408456 [astro-ph]}].
 
\bibitem{Takada:2006xs}
M.~Takada,
Phys. Rev. D \textbf{74} (2006), 043505
[\href{https://arxiv.org/abs/astro-ph/0606533}{[arXiv:astro-ph/0606533 [astro-ph]}].
   
\bibitem{Mehrabi:2015hva}
  A.~Mehrabi, S.~Basilakos and F.~Pace,
  Mon.\ Not.\ Roy.\ Astron.\ Soc.\  {\bf 452} (2015) no.3,  2930
 [\href{https://arxiv.org/abs/1504.01262}{arXiv:1504.01262 [astro-ph.CO]}].
 
 




\bibitem{TorresRodriguez:2007mk}
A.~Torres-Rodriguez and C.~M.~Cress,
Mon. Not. Roy. Astron. Soc. \textbf{376} (2007), 1831-1837
[\href{https://arxiv.org/abs/astro-ph/0702113}{arXiv:astro-ph/0702113 [astro-ph]}].
 
\bibitem{TorresRodriguez:2008et}
A.~Torres-Rodriguez, C.~M.~Cress and K.~Moodley,
Mon. Not. Roy. Astron. Soc. \textbf{388} (2008), 669
[\href{https://arxiv.org/abs/0804.2344}{arXiv:0804.2344 [astro-ph]}].
 
 
\bibitem{Sapone:2013wda}
  D.~Sapone, E.~Majerotto, M.~Kunz and B.~Garilli,
  Phys.\ Rev.\ D {\bf 88} (2013) 043503
 [\href{https://arxiv.org/abs/1305.1942}{arXiv:1305.1942 [astro-ph.CO]}].
 
 
 
\bibitem{Morais:2015ooa}
 J.~Morais, M.~Bouhmadi-L{\'o}pez and S.~Capozziello,
 JCAP {\bf 1509} (2015) no.09, 041
 [\href{http://arxiv.org/abs/1507.02623}{arXiv:1507.02623 [gr-qc]}].

\bibitem{Albarran:2016mdu}
  I.~Albarran, M.~Bouhmadi-L\'{o}pez and J.~Morais,
  Phys.\ Dark Univ.\  {\bf 16} (2017) 94
  [\href{https://arxiv.org/abs/arXiv:1611.00392}{arXiv:1611.00392 [astro-ph.CO]}].
  

 
    
 \bibitem{wikiesa}
\textit{Planck} 2018 Results: Cosmological Parameter Tables
 \href{ https://wiki.cosmos.esa.int/planck-legacy-archive/images/b/be/Baseline_params_table_2018_68pc.pdf}{ https://wiki.cosmos.esa.int}
 
 \bibitem{Bouali:2019whr}
A.~Bouali, I.~Albarran, M.~Bouhmadi-López, and T.~Ouali, ``{Cosmological
  constraints of phantom dark energy models},''
  \href{http://dx.doi.org/10.1016/j.dark.2019.100391}{{\em Phys. Dark Univ.}
  {\bfseries 26} (2019) 100391},
\href{http://arxiv.org/abs/1905.07304}{{\ttfamily arXiv:1905.07304
  [astro-ph.CO]}}.
 

 
\bibitem{Albarran:2017kzf}
  I.~Albarran, M.~Bouhmadi-López and J.~Morais,
  Eur.\ Phys.\ J.\ C {\bf 78} (2018) no.3,  260
   [\href{https://arxiv.org/abs/1706.01484}{[arXiv:1706.01484 [gr-qc]]}].
   
\bibitem{Abramowitz}
M.~{Abramowitz} and I.~A. {Stegun}, {\em {Handbook of mathematical functions
  with formulas, graphs, and mathematical tables}}.
\newblock 1965.



 
\end{thebibliography}
\end{document}